\documentclass[journal=ancac3,manuscript=article,layout=twocolumn]{achemso}
\setkeys{acs}{articletitle = true} % Will suppress showing article title

\UseRawInputEncoding
\usepackage[utf8]{inputenc}
\usepackage{array}
\usepackage{stfloats}
\usepackage{comment}
\usepackage{chemformula} % Formula subscripts using \ch{}
\usepackage[T1]{fontenc} % Use modern font encodings
\usepackage{float}
\usepackage{seqsplit} % force to split the line
\usepackage{graphicx}
\usepackage{booktabs} % table
\usepackage{multirow}
\usepackage{multicol}
\usepackage{color, colortbl}
\usepackage[para,flushleft]{threeparttable}
\usepackage{ulem}
\usepackage[super]{nth}
\usepackage{amsmath}
\usepackage[font=scriptsize,labelfont=bf]{caption}
\usepackage[switch]{lineno}
\usepackage{breqn}
\usepackage{adjustbox}
\usepackage{makecell} % Line break for table
\DeclareUnicodeCharacter{2009}{\,} 
\DeclareUnicodeCharacter{00C9}{\'E}
\DeclareUnicodeCharacter{00FC}{\"u}
\usepackage{chapterbib}
\mciteErrorOnUnknownfalse
\usepackage{amssymb}
\usepackage[version=4]{mhchem}
\usepackage{subcaption}
\usepackage{bbm}
\usepackage[shortlabels]{enumitem}
\usepackage{wrapfig}

\usepackage{tocloft}

\newcommand{\listofSIcontentsname}{Contents}
\newlistof{SIcontents}{sictoc}{\listofSIcontentsname}

\makeatletter
\renewcommand{\listofSIcontents}{%
  \section*{\small\bfseries \listofSIcontentsname}%
  \@starttoc{sictoc}%
}
\makeatother

\usepackage{etoolbox} % for \patchcmd
\makeatletter
\renewcommand*{\acs@author@fnsymbol@symbol}[1]{%
  \ifcase#1 *\or 1\or 2\or 3\or 4\or 5\or 6\or 7\or 8\or 9\or 10\fi
}

\patchcmd{\acs@address@list@auxii}
  {\acs@author@fnsymbol{\acs@affil@marker@cnt}}
  {\textsuperscript{\acs@author@fnsymbol{\acs@affil@marker@cnt}}}
  {}{}

\patchcmd{\acs@address@list@auxii}
  {{\acs@author@fnsymbol{\acs@affil@marker@cnt}\@nameuse{@altaffil@\@roman\@tempcnta}\par}}
  {{\textsuperscript{\acs@author@fnsymbol{\acs@affil@marker@cnt}}\@nameuse{@altaffil@\@roman\@tempcnta}\par}}
  {}{}
\makeatother
\usepackage{titletoc} 
\author{Binay P. Nayak}
\affiliation{Department of Chemical and Biological Engineering, Iowa State University, Ames, IA, 50011, US}
\alsoaffiliation{Division of Materials Sciences and Engineering, Ames National Laboratory, Ames, IA, 50011, US}

\author{Prapti Kakkar}
\affiliation{Department of Chemical and Biological Engineering, Iowa State University, Ames, IA, 50011, US}
\alsoaffiliation{Division of Materials Sciences and Engineering, Ames National Laboratory, Ames, IA, 50011, US}

\author{Wesley P. Korba}
\affiliation{Division of Materials Sciences and Engineering, Ames National Laboratory, Ames, IA, 50011, US}
\alsoaffiliation{Department of Physics and Astronomy, Iowa State University, Ames, IA, 50011, US}

\author{Honghu Zhang}
\affiliation{National Synchrotron Light Source II, Brookhaven National Laboratory, Upton, NY, 11973, US}

\author{Wenjie Wang}
\affiliation{Division of Materials Sciences and Engineering, Ames National Laboratory, Ames, IA, 50011, US}

\author{Surya K. Mallapragada}
\email{suryakm@iastate.edu}
\affiliation{Department of Chemical and Biological Engineering, Iowa State University, Ames, IA, 50011, US}
\alsoaffiliation{Division of Materials Sciences and Engineering, Ames National Laboratory, Ames, IA, 50011, US}
 
\author{Alex Travesset}
\email{trvsst@ameslab.gov}
\affiliation{Division of Materials Sciences and Engineering, Ames National Laboratory, Ames, IA, 50011, US}
\alsoaffiliation{Department of Physics and Astronomy, Iowa State University, Ames, IA, 50011, US}

\author{David Vaknin}
\email{vaknin@ameslab.gov}
\affiliation{Division of Materials Sciences and Engineering, Ames National Laboratory, Ames, IA, 50011, US}
\alsoaffiliation{Department of Physics and Astronomy, Iowa State University, Ames, IA, 50011, US}

\title{\Large Electrostatic Superlattices beyond 1:1 Stoichiometry}

\keywords{Nanoparticle self-assembly, ionic crystal analogs, photonic superlattices, small-angle X-ray scattering, gold nanoparticles}

\begin{document}

\begin{abstract}

Exotic nanoparticle superstructures can be accessed by harnessing nanoparticle softness and charge regulation, features often viewed as obstacles to structural control. Here, we show that regulated charge mismatch in polymer-grafted nanoparticles enables the assembly of high-stoichiometry cubic superlattices. By co-tuning grafting density, particle size, and bulk composition, we realize ionic-lattice analogues such as CaF$_2$ and Th$_3$P$_4$, as well as single-component A$_3$ and A$_7$ superlattices without atomic counterparts. The A$_3$ lattice has recently been identified theoretically as a photonic band-gap lattice. These phases emerge from a 1:1 “parent” lattice when local charge neutrality cannot be satisfied, driving either progressive interstitial filling or reorganization into a larger basis. For instance, the systematic occupation of ZnS tetrahedral sites yields CaF$_2$, while ligand-swapping symmetry breaking converts CsCl into Th$_3$P$_4$. Upon heating, the assemblies exhibit reversible lattice contraction and pronounced negative thermal expansion. Furthermore, the energetic penalty for defects increases with nanoparticle size, facilitating the scalable production of high-quality, open superlattices for photonic applications.

\end{abstract}

\section*{Introduction}

Binary ionic compounds crystallize into a diverse family of cubic structures governed primarily by valence and ionic size\cite{Pauling1929,West2014solid}. Complex stoichiometries often emerge from simpler 1:1 ($AB$) parent frameworks through well-defined structural hierarchies. For example, ZnS evolves into fluorite (\ch{CaF2}) through progressive occupation of tetrahedral interstitial sites, while CsCl-type (pseudo-bcc) lattice reorganizes into the clustered bcc-like \ch{Th3P4} phase with a 28-particle unit cell\cite{Thaner2016,Wang2017c,Hyde1989inorganic}. Translating such hierarchical, non-close-packed motifs to nanoparticle (NP) superlattices is challenging because NP building blocks are large, deformable, and lack fixed charges.

% These intrinsically open structures are stabilized only within narrow ionic size ratios, producing orthogonal or tetrahedral coordination motifs rather than maximally dense triangular packings. 

\begin{figure*}[!hbt]
 	\centering 
 	\includegraphics[width=0.8 \linewidth]{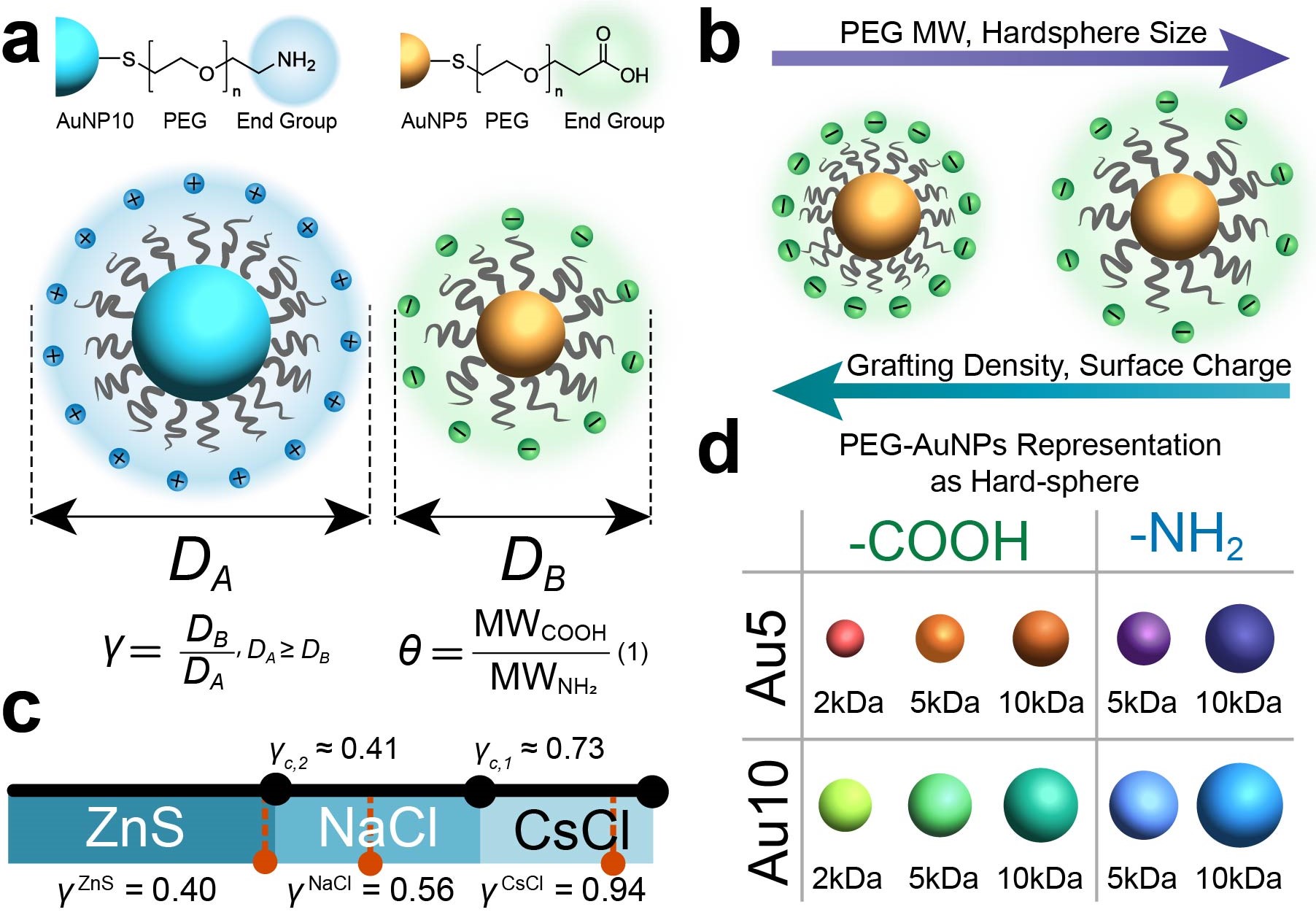}
   \caption{\textbf{Charge-regulated nanoparticle building blocks for superlattices beyond 1:1 stoichiometry.}
    (a) Schematic of gold nanoparticles (AuNPs) with core diameters $D_{\rm core}=5$~nm (yellow) and 10~nm (blue) functionalized with thiol-terminated poly(ethylene glycol) (PEG) bearing either \ch{-NH2} (blue, $+$) or \ch{-COOH} (green, $-$) end groups; the sign of the corona charge is indicated throughout. The polymer-grafted particles are treated as effective hard spheres of diameters $D_A$ and $D_B$, defining the size ratio $\gamma=D_B/D_A$, while $\theta$ denotes the molecular weight (MW) ratio of the oppositely charged PEG coronas. For identical PEG MW, the smaller, higher-curvature cores typically graft at a higher areal density, whereas the larger cores carry a larger total number of ligands per NP, as their surface area is greater; as a result, surface charge density and total particle charge need not scale in the same way.
    (b) Effect of PEG MW on effective particle size and grafting: increasing PEG MW increases the hard-sphere diameter $D_{\rm H}$, while reducing grafting density and the resulting surface charge density on the NP. 
    (c) Schematic structural sequence expected for binary $AB$ crystals as a function of the hard-sphere size ratio $\gamma$: ZnS ($\gamma<\gamma_{c,2}$) $\rightarrow$ NaCl ($\gamma_{c,2}<\gamma<\gamma_{c,1}$) $\rightarrow$ CsCl ($\gamma>\gamma_{c,1}$), where $\gamma_{c,2}=\sqrt{2}-1\simeq 0.414$ and $\gamma_{c,1}=\sqrt{3}-1\simeq 0.732$. $\gamma_{c}$ defined as critical $\gamma$ required to form a ionic lattice. Representative atomic analogues are shown for reference: Cs$^+$Cl$^-$ ($\gamma=0.94$) adopts CsCl symmetry, while Na$^+$Cl$^-$ ($\gamma=0.56$) and K$^+$Cl$^-$ ($\gamma=0.73$) crystallize in the rock-salt (NaCl) structure. For divalent systems, larger size asymmetries are typical; Zn$^{2+}$S$^{2-}$ ($\gamma\approx 0.40$) forms zinc-blende (fcc) or wurtzite (hcp) depending on external conditions such as pressure and temperature. Dashed red lines denote the ideal theoretical boundary values for each structural regime.
    (d) Color key and schematic effective hard-sphere sizes used throughout this work; darker shades correspond to larger PEG MW. Full NP nomenclature is provided in Table~S1.
    }
\vspace{-0.3 cm}
\label{fig:schematics} 
 \end{figure*}

DNA-mediated assembly realizes such superlattices by programming NP interactions via complementary DNA base pairing, allowing access to a broad range of lattice symmetries, including open superstructures \cite{Mirkin1996,Alivisatos1996,nykypanchuk2008dna,auyeung2012synthetically,macfarlane2011nanoparticle}. DNA origami further extends this concept by providing rigid templates that define NP positions with high precision \cite{Rothemund2006,Liu2016a,tian2016lattice,Liu2024theory}. Despite their versatility, these methods may face practical constraints, including multi-step component preparation, material costs, and challenges in scaling up \cite{seeman2017dna}. In contrast, solvent-evaporation-driven assembly of short-ligand-coated NPs offers scalability and simplicity, producing numerous binary superlattices \cite{Whetten1996,Murray1995,Shevchenko2006,Talapin2009,Boles2016}. However, the dominance of short-chain ligands in evaporation-driven assembly biases crystallization toward close-packed lattices, leaving open structures largely inaccessible \cite{travesset_nanocrystal_2024,Bassani2024}.

Electrostatic assembly offers a complementary route to mimicking binary ionic crystals by combining the scalability of solvent-based approaches with the ability to stabilize open, non-close-packed structures. Early demonstrations at the micrometer scale revealed a rich set of ionic analogs, including NaCl, CsCl, and large-small(LS)-type lattices \cite{Leunissen2005}, while subsequent studies extended electrostatic design principles to submicrometer and nanoscale systems \cite{kostiainen_electrostatic_2013,kalsin2006electrostatic,bian2021electrostatic}. These studies established a foundation for charge-driven nanoparticle assembly at the nanoscale, demonstrating that electrostatic interactions can organize diverse building blocks into ordered structures. Yet, many nanoscale electrostatic assemblies remain limited to close-packed, cluster-like or short-range ordered structures, and a general route to solution-state crystalline superlattices spanning multiple open symmetries, particularly beyond 1:1 stoichiometry, remains underdeveloped \cite{wang2024charge,hagan_equilibrium_2021,morozova_colloidal_2023}. Building on these foundations, recent advances have shown that polymer-functionalized NPs can act as effective ionic building blocks, enabling binary superlattices whose phase diagrams are governed by two parameters 
\begin{equation}
    \gamma = \frac{D_B}{D_A} \quad , \quad \theta = \frac{(\mbox{MW})_{\rm COOH}}{(\mbox{MW})_{\rm NH_2}}
\end{equation}
where $D_A$ and $D_B$ are the effective hard-sphere (HS) diameters of the larger $A$ and smaller $B$ NPs, respectively \cite{Zha2020,travesset_nanocrystal_2024}, and $\mathrm{MW}$ is the molecular weight of the grafted polymer. In the experimental systems we consider here, HS diameters are consistent with the hydrodynamic diameters ($D_{\mathrm{H}}$) of NPs. As the size ratio $\gamma$ varies, binary ionic systems follow a well-known structural sequence across the ZnS–NaCl–CsCl family, as summarized in this trend and representative ionic examples in Fig.~\ref{fig:schematics}. For NP systems, the critical size ratio $\gamma_c$ acquires an additional, monotonic dependence on the charge ratio $\theta$ arising from electrostatic correlations. These correlations further stabilize open lattices such as ZnS (and, to a lesser extent, NaCl), effectively shifting $\gamma_c$ to larger values as $\theta$ increases \cite{nayak2026valence}.

% As a function of $\gamma$, a specific sequence of phases is 
% \begin{equation}\label{Eq:phases}
%     \mbox{ZnS} \stackrel{\gamma_c=\sqrt{2}-1}{\rightarrow} \mbox{NaCl} \stackrel{\gamma_c=\sqrt{3}-1}{\rightarrow} \mbox{CsCl} 
% \end{equation}
% The phase diagram implied by Eq.~\ref{Eq:phases}\cite{nayak2026valence} captures the structural trends of simple binary ionic salts with 1:1 stoichiometry. For instance, Cs$^+$Cl$^-$ ($\gamma = 0.94$) crystallizes in the CsCl structure, whereas systems with moderate size asymmetry, such as Na$^+$Cl$^-$ ($\gamma = 0.56$) and K$^+$Cl$^-$ ($\gamma = 0.73$), adopt the rock-salt structure. For divalent compounds, larger size asymmetries are stabilized: Zn$^{2+}$S$^{2-}$ ($\gamma = 0.41$) crystallizes in either the zinc-blende (FCC) or wurtzite (hexagonal) lattice, depending on external conditions such as pressure and temperature \cite{desgreniers_pressure-induced_2000}. 

In simple ionic crystals, charge neutrality is intrinsic. By contrast, in aqueous suspensions, NP assemblies can have their effective charges screened or balanced by mobile counterions in the presence of added electrolytes. Recent results \cite{nayak2026valence} indicate that NP charges can self-adjust to satisfy an overall neutrality condition, which we term ``nanoparticle neutrality''. Under this condition, counterions attached to the grafted polymers are released into the surrounding solution for an entropic gain, thus facilitating assembly and crystallization \cite{Travesset2024a}. General phase diagrams for charged components in the presence of counterions have been developed previously \cite{hynninen_prediction_2006}; when adapted for NP neutrality, these predict stable superlattices beyond 1:1 stoichiometry, including $A_mB_n$ phases such as \ch{CaF2} (fluorite) and Cr$_3$Si (Frank--Kasper A15). 

This study focuses on extending electrostatic NP assembly from previously demonstrated ordered or short-range ordered nanoscale structures toward solution-state crystalline superlattices with general stoichiometry. To access these phases, we introduce grafting asymmetry between the two NP species, biasing charge regulation away from 1:1 balance and providing a controllable route to higher-order open superlattices \cite{Travesset2024a}.
This is achieved by tuning the grafting density of the polymer coronas through two independent strategies: (i) exploiting molecular-weight–dependent grafting effects (ii) varying the NP core curvature, which has been shown experimentally and theoretically to strongly affect the grafting density of polymer chains\cite{milner1991polymer}, as we have quantified in previous studies\cite{KimTravesset2021,nayak2025effect}. Experimentally, we implement this design by functionalizing gold nanoparticles (AuNPs) with thermo- and pH-responsive water-soluble polymers terminated with thiols at one end and either \ch{NH2} or \ch{COOH} groups at the other, enabling independent control of surface charge and polymer conformation. Using \textit{in situ} small-angle X-ray scattering (SAXS), we investigate how charge mismatch, composition, and temperature collectively direct the evolution of cubic superlattices. Our goal is to identify conditions under which open ZnS/diamond-type motifs progress toward \ch{CaF2} (a superlattice previously not reported in the NP assembly literature), and, likewise, when CsCl-derived frameworks reorganize into \ch{Th3P4}-type superlattices, thus establishing progressive site distribution as a general route to higher-stoichiometry NP superlattices.

\section{Results}
To translate the structural hierarchy into experiment, we tune the effective size ratio and charge asymmetry of polymer-grafted NPs to drive transitions from parent $AB$ lattices toward higher-stoichiometry $A_mB_n$  phases. Beginning with ZnS-type assemblies exhibiting partial interstitial occupation, we induce complete site filling to form \ch{CaF2}, and subsequently access the clustered bcc-like \ch{Th3P4} superstructure through ligand-driven symmetry breaking.

\begin{figure}[!hbt]
 	\centering 
 	\includegraphics[width=1\linewidth]{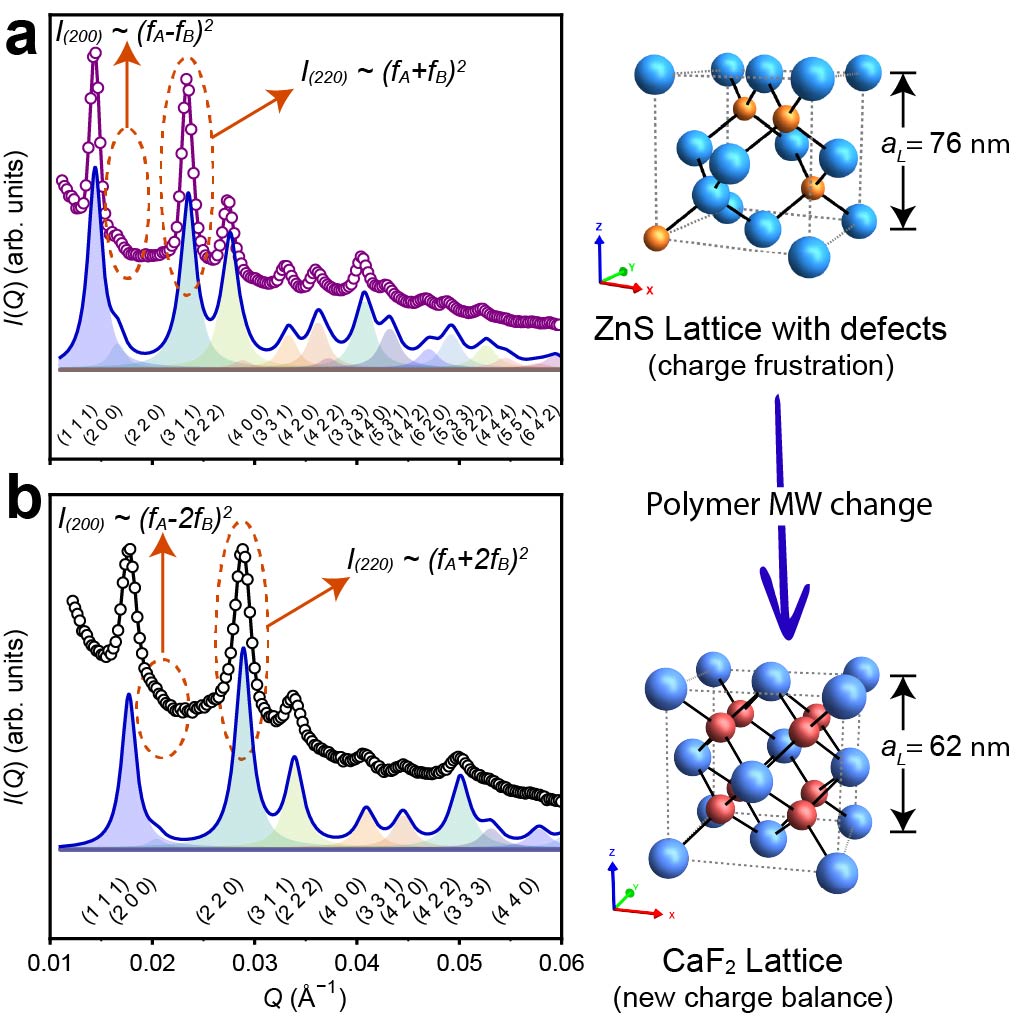}
    \caption{\textbf{Molecular weight induced grafting density transition from ZnS to \ch{CaF2}.}
    (a) Small-angle X-ray scattering (SAXS) intensity, $I(Q)$ vs.\ $Q$, for COOH--PEG5k--Au5 and \ch{NH2}--PEG10k--Au10 NPs mixed at a 2:1 number ratio. In a previous study, we demonstrated that a size ratio $\gamma \approx 0.7$ (defined as the ratio of the larger to smaller functionalized AuNP diameters) stabilizes a defect-rich ZnS-type superlattice. In this system, the ZnS structure is formed by COOH--PEG5k--Au5 and \ch{NH2}--PEG10k--Au10, and its pronounced defectivity is attributed to an imbalance in surface charge arising from asymmetric polymer coronas (see discussion below), leading to interstitial occupation or site-exchange defects. To test this hypothesis, we reduce the grafted PEG molecular weights on both particle species, replacing the coronas with PEG2k--COOH and PEG5k--\ch{NH2}. 
    (b) SAXS intensity for COOH--PEG2k--Au5 and \ch{NH2}--PEG5k--Au10 mixed at the same 2:1 number ratio, showing crystallization into a well-ordered \ch{CaF2}-like superlattice at room temperature. Experimental data are shown as open circles, while modeled intensity profiles are plotted as solid blue lines, with individual Bragg peak contributions shaded by color. Corresponding Miller indices are indicated below each shaded contribution. Schematic illustrations of the defect-rich ZnS lattice and the \ch{CaF2} superstructure are shown adjacent to their respective diffraction patterns. All SAXS profiles are individually normalized and displayed on a linear intensity scale. Dotted red ovals highlight the (200) and (220) Miller reflections, whose relative intensities arise from constructive and destructive interference.
}
    \vspace{-0.3 cm}
\label{fig:caf2vszns} 
 \end{figure}

\subsection{Grafting-Controlled Transition from ZnS to \ch{CaF2} Superlattices}

Figure~\ref{fig:caf2vszns} summarizes the structural evolution obtained by systematically varying the surface chemistry of the binary NP system while keeping the core sizes fixed (10 and 5 nm). For COOH–PEG5k–Au5 and \ch{NH2}–PEG10k–Au10 mixed at a 2:1 number ratio, the SAXS pattern (Fig.~\ref{fig:caf2vszns}a) is well described by a ZnS-type superlattice, albeit with signatures of nearest-neighbor site-exchange defects. Consistent with our previous results, a size ratio $\gamma \approx 0.7$ stabilizes the ZnS framework; however, the relative peak intensities and diffuse background reveal interstitial occupation and local disorder rather than an ideal ZnS lattice. While changing the PEG MW modifies both the effective size ratio and the corona thickness, the resulting systems remain within the $\gamma$ range that favors a ZnS parent lattice. We therefore attribute the defect-rich character primarily to asymmetric grafting densities, which generate an imbalance in regulated surface charge and frustrate strict 1:1 charge neutrality within the ZnS unit cell.\cite{nayak2026valence} In this picture, $\gamma$ determines the parent lattice symmetry, whereas charge asymmetry controls the extent of interstitial occupation and the transition toward higher-stoichiometry derivatives.
Furthermore, 2:1 mixture of COOH–PEG2k–Au5 and \ch{NH2}–PEG10k–Au10 exhibits the same interstitial-defect-rich ZnS structure (see Supplementary Information (SI) , Fig. S7). To test whether controlled modification of grafting density could promote more complete occupation of interstitial sites, we replaced the PEG5k/PEG10k coronas with shorter PEG2k/PEG5k chains while maintaining the same NP core sizes and overall mixing ratio. The resulting SAXS pattern (Fig.~\ref{fig:caf2vszns}b) exhibits a clear enhancement of long-range order (LRO) and is well described by a polycrystalline \ch{CaF2}-type structural model. This transition is accompanied by changes in both the effective size ratio and the regulated charge asymmetry; however, unlike the parent ZnS framework, stabilization of the \ch{CaF2} phase requires substantial interstitial occupation, which we associate primarily with the increased charge asymmetry introduced by the modified grafting densities. Although the 2:1 mixture still shows a non-negligible diffuse background, reflecting residual disorder that can be further reduced by increasing the concentration of the 5~nm component in the bulk, as discussed in a subsequent section. The emergence of the full set of fluorite Bragg reflections and the quality of the structure-factor fit indicate that the system closely approaches a stoichiometric \ch{CaF2}-like superlattice at room temperature. The identification of the underlying superlattice does not rely solely on quantitative fitting; instead, distinct qualitative features in the X-ray diffraction profiles provide robust structural fingerprints as discussed in detail in the SI Section~4.9.

\subsection{Complementary Stoichiometric Control through Bulk Mixing.}

% Having established that the \ch{CaF2} superlattice can be accessed through controlled modification of a ZnS framework by tuning the polymer molecular weights (Fig.~\ref{fig:caf2vszns}), we now explore a complementary and more direct route based solely on controlling the bulk number ratio of the two nanoparticle species. In these experiments, we fix the surface chemistry to the same molecular weight combination that stabilizes the \ch{CaF2} phase in Fig.~\ref{fig:caf2vszns}b and systematically vary the mixing ratio in solution.

\begin{figure*}[!hbt]
 	\centering 
 	\includegraphics[width=1\linewidth]{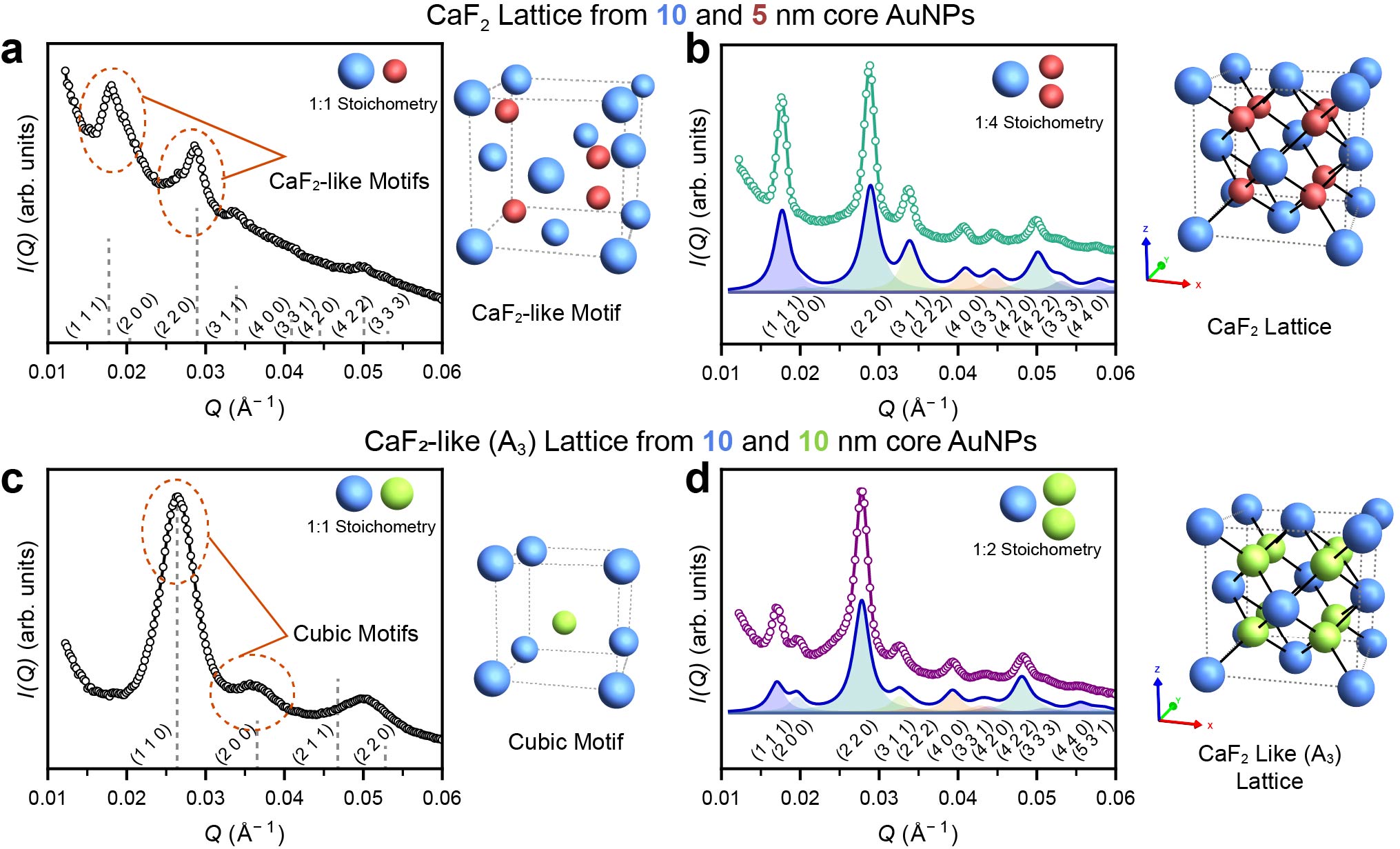}
   \caption{\textbf{Emergence of a stoichiometric \ch{CaF2} superlattice from bulk mixing ratio.}
(a) SAXS diffraction pattern, $I(Q)$ vs.\ $Q$, for COOH--PEG2k--Au5 and \ch{NH2}--PEG5k--Au10 mixed at a 1:1 number ratio. Although this composition was predicted to favor a \ch{CaF2}-like structure, the pattern exhibits only short-range order (SRO), indicating that the exact \ch{CaF2} stoichiometry is highly sensitive to the bulk particle ratio.
(b) Increasing the mixing ratio 4:1 drives crystallization into a \ch{CaF2}-type superlattice, evidenced by well-defined Bragg reflections. The SAXS profile is well described by a polycrystalline \ch{CaF2}-like model. At a 2:1 ratio, the structure is of high quality, but a residual background persists, consistent with inhomogeneous charge fluctuations on the smaller particles, such that only a fraction of the population satisfies the crystallization condition.
(c) SAXS pattern for a binary system with identical core sizes (10~nm) and the same functional groups as in (a), mixed at a 1:1 ratio. Only the SRO characteristic of a cubic-like structure is observed, yielding a limited number of broadened Bragg reflections.
(d) For the same 10~nm cores at a 2:1 ratio, the system crystallizes into a highly ordered \ch{CaF2} superlattice. In contrast to (a) and (c), the PEG corona conformation and charge distribution are fully compatible with \ch{CaF2} formation at the matched stoichiometry in the suspension. Here, the polymeric functionalization acts as a soft ``framework'' that defines the available coordination sites for the NPs; consequently, the NP positions become the primary signatures of the emergent lattice symmetry. In this sense, the resulting \ch{CaF2} superlattice can be regarded as an \ch{A3}-type formal stoichiometry for the NPs, with no direct atomic analog in nature. Experimental data are shown as open circles, while modeled intensity profiles are plotted as solid blue lines, with individual Bragg peak contributions shaded by color. Corresponding Miller indices are indicated below each shaded contribution. Schematic illustrations of the SRO structures and the \ch{CaF2} superstructure are shown adjacent to their respective diffraction patterns. All SAXS profiles are individually normalized and displayed on a linear intensity scale.}
    \vspace{-0.3 cm}
\label{fig:caf2_main} 
 \end{figure*}
 
Next, we test a composition-driven route to achieve the \ch{CaF2} structure. Holding the PEG2k/PEG5k functionalization fixed, we systematically vary the bulk number ratio of the two NP species to determine how stoichiometry alone selects the superlattice.
Accordingly, Fig.~\ref{fig:caf2_main}a displays only broad diffraction features for the 1:1 mixture, indicative of short-range order (SRO). Notably, these diffuse peaks occur at positions corresponding to the \ch{CaF2} reflections, suggesting the presence of locally ordered \ch{CaF2}-like motifs that fail to propagate into long-range crystalline order.

When the bulk mixing ratio is adjusted to match the \ch{CaF2} stoichiometry, long-range crystallization emerges. Figure~\ref{fig:caf2_main}b shows well-defined Bragg reflections characteristic of a \ch{CaF2} superlattice. While a 2:1 mixing ratio already stabilizes the fluorite structure, increasing the bulk ratio to 4:1 significantly suppresses the diffuse background, yielding improved long-range order and enhanced crystallinity. This confirms that the 1:2 \ch{CaF2} lattice is the thermodynamically selected crystalline phase under these conditions. We attribute this improvement to selective incorporation: increasing the concentration of the smaller component provides a larger pool of particles whose size, grafting density, and regulated charge are compatible with the \ch{CaF2} lattice. Excess particles are excluded from the ordered domains, as indicated by the absence of additional Bragg reflections or defect-superstructure peaks and by residual individual-particle form-factor scattering.

\begin{figure}[H]
 	\centering 
 	\includegraphics[width=1 \linewidth]{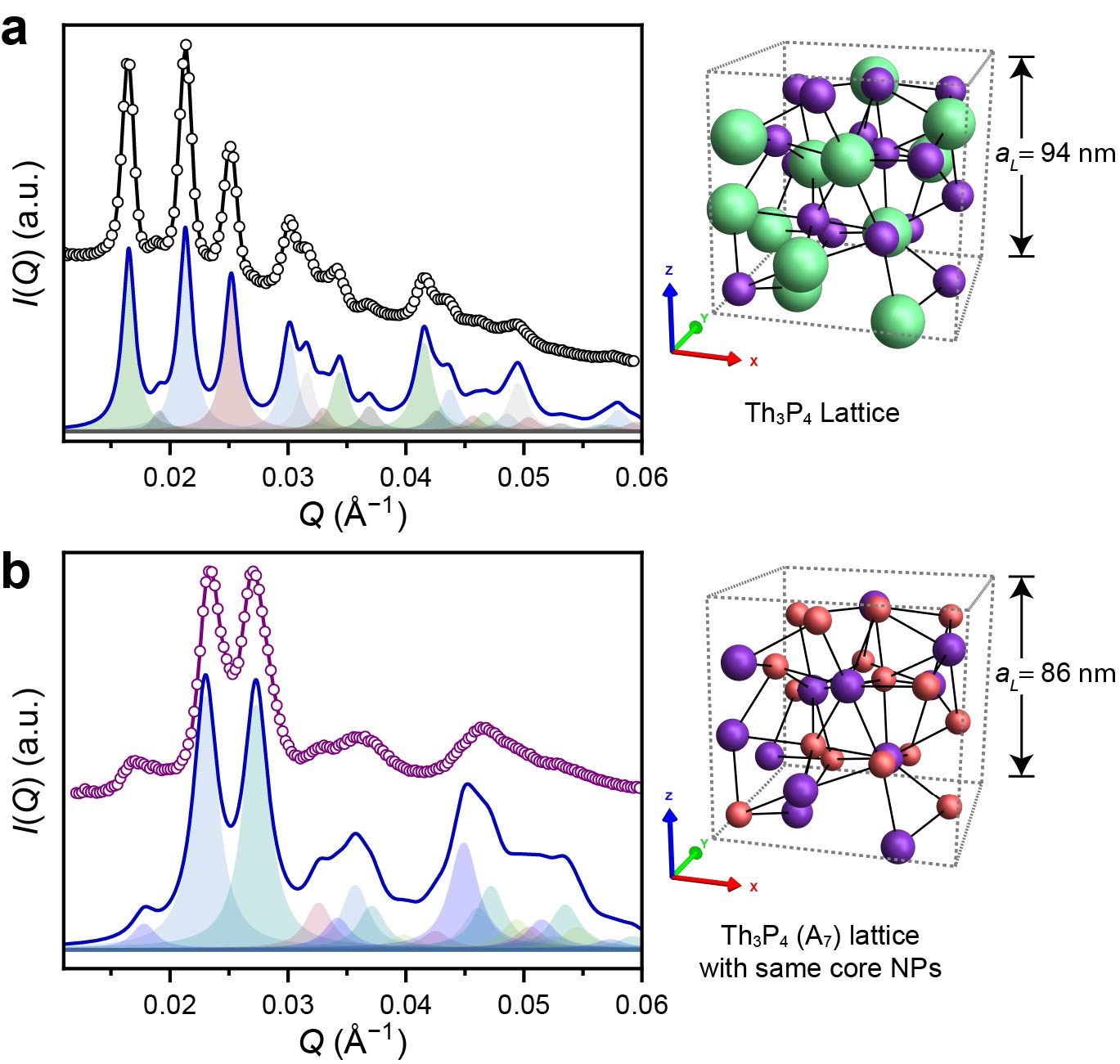}
   \caption{\textbf{Symmetry breaking from CsCl to higher-stoichiometry \ch{Th3P4} superlattices.}
    (a) SAXS intensity, $I(Q)$ vs.\ $Q$, for a 4:1 number-ratio mixture of \ch{NH2}--PEG5k--Au5 and \ch{COOH}--PEG5k--Au10 at pH~3. The diffraction pattern is well described by an ideal \ch{Th3P4}-type model. For the inverse ligand assignment (\ch{COOH}--PEG5k--Au5 with \ch{NH2}--PEG5k--Au10), the same core-size pair forms a CsCl (\ch{$AB$}) superlattice~\cite{nayak2026valence}. Swapping the ligand identities changes the curvature-dependent grafting density and hence the regulated surface charge densities, breaking the effective 1:1 charge balance that stabilizes CsCl and instead favoring a higher-stoichiometry \ch{Th3P4} superstructure (28 NPs per unit cell; nominal $A:B=3:4$).
    (b) SAXS intensity for COOH--PEG2k--Au5 and \ch{NH2}--PEG5k--Au5 mixed at a 1:1 number ratio. Despite identical core sizes, asymmetric polymer coronas generate an effective size/charge asymmetry that stabilizes a meso-ordered \ch{Th3P4}-like superlattice. Since the polymer corona can be treated as a framework, the effective lattice is equivalent to a \ch{A7} lattice without any natural analogue. Experimental data are shown as open circles; modeled intensity profiles are shown as solid lines with individual Bragg contributions shaded. Schematic unit-cell illustrations are shown alongside the corresponding SAXS patterns. All datasets are normalized individually and plotted on a linear intensity scale.}
    \vspace{-0.3 cm}
\label{fig:th3p4} 
 \end{figure}
 
We further extend this concept by considering a binary system composed of equal-sized NPs (10~nm cores), which corresponds to a hypothetical \ch{A3} stoichiometry with no direct atomic analogue. At a 1:1 mixing ratio, the SAXS pattern (Fig.~\ref{fig:caf2_main}c) exhibits only broadened peaks, consistent with short-range cubic motifs and the absence of long-range order. Notably, when the mixing ratio is adjusted to 2:1, thereby matching the effective \ch{A3} stoichiometry, the system crystallizes into a highly ordered pseudo-\ch{CaF2} superlattice as shown in Fig.~\ref{fig:caf2_main}d. The 4:1 mixture under similar conditions crystallizes in the same structure \ch{A3} as shown in Fig.~S6.

In the equal-sphere description adopted here, this lattice is geometrically equivalent to the \ch{Li2O} framework, because fluorite and anti-fluorite structures differ only by interchange of the occupied sublattices. We therefore refer to the resulting architecture generically as a \ch{A3} lattice of identical spheres. Within this representation, the corresponding \ch{A3} lattice has been shown to support complete photonic band gaps over selected filling-fraction and dielectric-contrast ranges.\cite{cersonsky2021diversity}

% This result highlights that, in nanoparticle assemblies, polymer-mediated charge regulation enables access to stoichiometries and lattice symmetries that have no counterpart in atomic crystals. The \ch{A3} lattice has an estimated packing fraction of $\sim$51\%, intermediate between simple-cubic ($\sim$52\%) and diamond ($\sim$33\%) lattices, both of which are known to exhibit photonic band-gap properties.

 \subsection{Ligand-swap Symmetry-breaking and access to \ch{Th3P4} Stoichiometry}

Figure~\ref{fig:th3p4} shows that small changes in surface chemistry can redirect assembly from a unit-stoichiometry CsCl lattice to a higher-stoichiometry \ch{Th3P4} superstructure. In our previous work, the COOH--PEG5k--Au5/\ch{NH2}--PEG5k--Au10 pair crystallized into CsCl, consistent with an effective 1:1 charge balance between the two components~\cite{nayak2026valence}. Notably, exchanging the ligand identities (\ch{NH2} on the 5~nm cores and COOH on the 10~nm cores) stabilizes a \ch{Th3P4}-type superlattice (Fig.~\ref{fig:th3p4}a). We attribute this transition to curvature-dependent changes in grafting density that modify the regulated surface charge densities, thereby breaking the near-unit charge symmetry required for CsCl and favoring a $3:4$ (nominal) occupation of the unit cell. 
The refined \ch{Th3P4}-type unit cell has a lattice constant of $\sim$94~nm and contains 28 NPs distributed over the crystallographic 12$a$ (Th) and 16$c$ (P) sites. A full structural discussion is provided in the SI Section~6. Despite the large lattice parameter, the nearest-neighbor distances remain $\sim$30 nm, comparable to CsCl assemblies, thus validating the concept of ``hard sphere diameter'' previously introduced and indicating that the expanded unit cell originates from symmetry and basis complexity rather than increased interparticle spacing. The 28-particle basis reflects a balance between short-range electrostatic correlations and steric packing constraints, leading to ordered \ch{Th3P4} domains with finite coherence length.

Our attempt to populate all 28 lattice sites of the \ch{Th3P4} structure with the same core NP (10 or 5 nm) without altering the grafted polymer yielded a more complex BCC-like superstructure that could not be readily assigned to a specific space group at this time. This behavior underscores the phase's sensitivity to the effective size ratio, $\gamma$, and to charge asymmetry. In practice, the ordering of \ch{Th3P4} can be recovered by tuning $\gamma$ through the MWs of PEG, which changes the effective hard-sphere size and the regulated surface-charge balance. An analogous route to \ch{Th3P4}-like meso-range ordering (MRO) is accessible even for equal core sizes: asymmetric coronas generate an effective size/charge asymmetry sufficient to stabilize the same symmetry as demonstrated in Fig.~\ref{fig:th3p4}(b), for  COOH--PEG2k--Au5 and \ch{NH2}--PEG5k--Au5. This structure is unique in crystallography, as each lattice site can be treated as a single atom, effectively translating the stoichiometry to A7, which lacks a parallel atomic analog. 

\begin{figure*}[!hbt]
 	\centering 
 	\includegraphics[width=.8\linewidth]{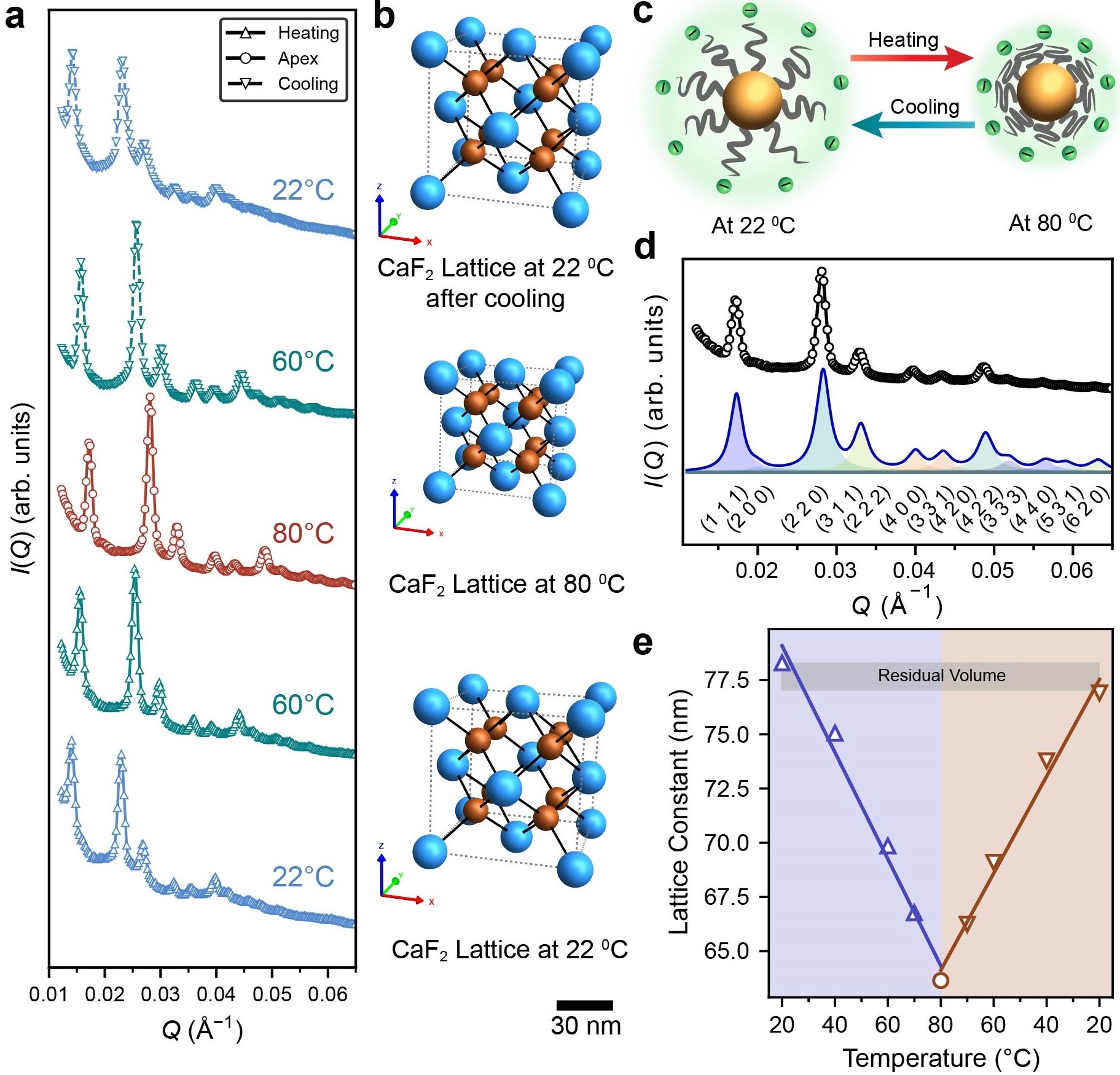}
    \caption{\textbf{Thermal stabilization of a \ch{CaF2} nanoparticle superlattice.}
    (a) SAXS diffraction patterns ($I(Q)$ vs.\ $Q$) for COOH--PEG10k--Au5 and \ch{NH2}--PEG10k--Au10 at pH~3 (2:1 number ratio) collected during a heating--cooling cycle. Each profile is normalized independently and shown on a linear scale. The diffraction patterns are vertically shifted for clarity. At room temperature (22~$^\circ$C), the mixture crystallizes into an imperfect \ch{CaF2} phase. Upon heating above 40~$^\circ$C, the diffraction peaks sharpen, and additional reflections emerge, indicating markedly improved crystallinity; subsequent cooling reversibly restores the room-temperature lattice quality.
    (b) Schematic of the temperature-dependent structural evolution. The high-temperature state is depicted with a reduced effective unit cell to reflect corona contraction, which correlates with enhanced ordering.
    (c) Schematic illustration of the grafted PEG corona in its swollen (room-temperature) state and its collapsed conformation upon heating to 80~$^\circ$C, highlighting the reduction in effective $D_{\rm H}$ at elevated temperature.
    (d) Representative SAXS pattern at 80~$^\circ$C overlaid with the simulated \ch{CaF2} structure-factor intensity, highlighting the superior crystal quality at elevated temperature. Experimental data are shown as open circles; modeled intensity profiles are shown as solid lines with individual Bragg contributions shaded. 
    (e) Extracted lattice parameter as a function of temperature, revealing reversible negative thermal expansion. The lattice constant decreases from $\sim$77.5~nm at 22~$^\circ$C to $\sim$62.9~nm at 80~$^\circ$C (a $\sim$20\% reduction in $a$), corresponding to an approximately two-fold decrease in unit-cell volume ($\Delta V/V \approx 1-(62.9/77.5)^3 \sim 0.47$), consistent with thermal contraction of the PEG corona.}
    \vspace{-0.3 cm}
\label{fig:caf2_temp} 
 \end{figure*}

\subsection{Thermal Effects and Negative Thermal Expansion in \ch{CaF2} Superlattices}

To probe the role of thermal fluctuations in soft superlattices, we investigate the reversible temperature-dependent reorganization of the \ch{CaF2} lattice, rather than irreversible annealing as in atomic crystals. Figure~\ref{fig:caf2_temp}a presents in situ SAXS measurements collected during a heating–cooling cycle for COOH--PEG10k--Au5 and \ch{NH2}--PEG10k--Au10 assembled at pH~3 and a 2:1 number ratio. At room temperature, the diffraction pattern is consistent with a \ch{CaF2}-type structure, but exhibits broadened Bragg reflections, indicative of finite domain size and residual disorder. Upon heating above 40~$^\circ$C, the diffraction peaks sharpen markedly and additional reflections become resolvable, signaling an enhancement of long-range crystalline order. Importantly, this improvement is fully reversible: cooling restores the original diffraction pattern, demonstrating that the structural reorganization is thermodynamically controlled rather than kinetically trapped.

The superlattice exhibits a large, fully reversible contraction of the unit cell upon heating (Fig.~\ref{fig:caf2_temp}e). The lattice parameter decreases from 77.5~nm at 22~$^\circ$C to 62.6~nm at 80~$^\circ$C, corresponding to a $\sim$20\% reduction in $a$ and nearly a 50\% decrease in unit-cell volume, while the lattice symmetry remains unchanged throughout the heating–cooling cycle. The heating and cooling branches show only a small hysteresis, which we attribute to finite polymer relaxation kinetics during solvent expulsion and re-swelling of the PEG corona. The recovery of the room-temperature Bragg pattern after cooling indicates that the thermal response is reversible over the cycle studied and does not produce irreversible degradation of the superlattice. Notably, the temperature dependence of the lattice constant is linear with no anomaly in the measured temperature range, consistent with the fact that PEG's lower-critical solution temperature is above 100 \textdegree{C}. Such pronounced negative thermal expansion (NTE) is uncommon in crystalline solids and reflects the soft character of the assembly. The effect arises from the thermal response of PEG: the polymer chains expel solvent and collapse, reducing the effective corona volume and enabling closer NP packing. An analogous response is observed for equal core sizes (10/10~nm) with identical polymer MWs (see SI, Fig.~S17). In some cases, this thermal effect modifies the $\gamma$ ratio to that measured for the individual particle from DLS and can lead to spontaneous crystallization at higher temperatures. \cite{kim2021effects,nayak2023tuning}

\subsection{Size-Scalable Stabilization of the ZnS Superlattice}

Having established that controlled grafting asymmetry enables access to ZnS and \ch{CaF2} superlattices at the 5–10~nm scale, we next examine the robustness and scalability of this assembly strategy by increasing the NP core sizes. Specifically, we consider a binary system composed of COOH--PEG5k--Au20 and \ch{NH2}--PEG10k--Au10 NPs, corresponding to a twofold increase in core diameter relative to the systems discussed in Fig. \ref{fig:caf2vszns}a, while preserving the same surface functionalization and mixing ratio.

\begin{figure}[!hbt]
 	\centering 
 	\includegraphics[width=1\linewidth]{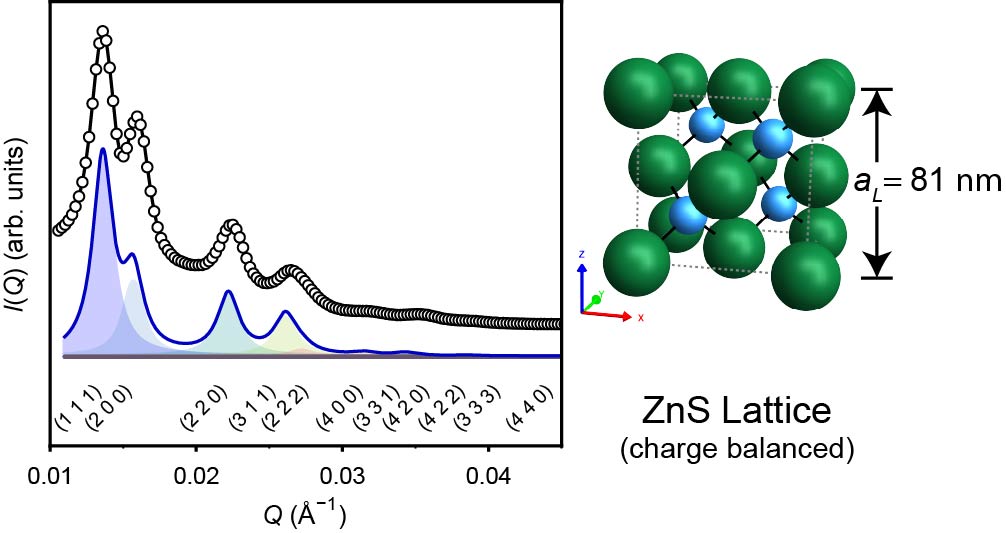}
      \caption{\textbf{Robust stabilization of the ZnS superlattice at larger nanoparticle sizes.} SAXS intensity, $I(Q)$ vs.\ $Q$, for COOH--PEG5k--Au20 and \ch{NH2}--PEG10k--Au10 NPs mixed at a 2:1 number ratio at pH~3 and room temperature. The surface functionalization mirrors that used in Fig.~\ref{fig:caf2vszns}a, but with NP cores scaled up by a factor of two (20 and 10~nm diameters). The 1:1 mixing ratio also forms the ZnS lattice as shown in the Supplementary Information (SI). The diffraction pattern exhibits well-defined Bragg reflections consistent with a highly ordered ZnS superlattice, indicating enhanced crystallinity relative to smaller-core systems. A detailed comparison between the experimental data and the expected ZnS structure factor is provided in the SI. Experimental data are shown as open circles; modeled intensity profiles are shown as solid lines with individual Bragg contributions shaded. A schematic illustration of the ZnS lattice with HS is shown next to the diffraction pattern. 
    }
    \vspace{-0.3 cm}
\label{fig:zns_20nm} 
 \end{figure}

As shown in Fig.~\ref{fig:zns_20nm}, the SAXS diffraction pattern displays well-defined Bragg reflections that are consistent with a highly ordered ZnS superlattice. Compared to the smaller-core analogue (Fig.~\ref{fig:caf2vszns}a), the larger NPs yield improved crystallinity with minimal diffuse background, indicating suppressed interstitial occupation and reduced site-exchange disorder. These results demonstrate that ZnS ordering is robust upon upscaling the NP size, and suggest that the associated changes in curvature-dependent grafting and charge distribution stabilize local charge balance within the ZnS framework without requiring interstitial incorporation. A similar size-stabilization trend is observed for NaCl superlattices (Fig.~S13), where increasing the NP core from 10~nm to 15~nm yields sharper reflections and reduced defect signatures, consistent with improved long-range order. These findings position NP size as a central design parameter for defect-free open superlattices, paving the way toward scalable diamond- and simple-cubic-type photonic materials.

Additional measurements presented in the SI further confirm the robustness of the ZnS phase under these conditions. For a 1:1 bulk mixing ratio of the same COOH--PEG5k--Au20 and \ch{NH2}--PEG10k--Au10 system, the diffraction patterns remain fully consistent with a stable ZnS superlattice. Moreover, attempts to drive crystallization toward a \ch{CaF2} phase by increasing the bulk mixing ratio to 4:1 were unsuccessful, underscoring the strong energetic preference for ZnS ordering at larger core sizes. 

% These observations highlight that, under this surface grafting and curvature regime, the ZnS lattice represents a particularly stable endpoint that resists transformation into higher-stoichiometry fluorite-type structures.

\section{Discussion}

Our results show that subtle control of grafting density and charge regulation enables access to a hierarchy of NP superlattices, all within a narrow pH window (3–4).\cite{nayak2026valence,nayak2023ionic,nayak2025electrostatically} In this regime, the effective charges of oppositely functionalized NP self-adjust toward near-neutrality, as previously demonstrated \cite{nayak2026valence}. To identify charge-neutral assembly conditions, we compare the measured $\zeta$-potentials of the two NP species. For 1:1 lattices, neutrality occurs where the positive and negative $\zeta$-potential magnitudes coincide; for higher stoichiometries, the curves are renormalized by the target particle ratio (Fig.~S26). The resulting intersections define the pH windows expected to stabilize each superlattice stoichiometry. We propose that this “NP neutrality” condition plays a central role in selecting superlattice stoichiometry: the regulated NP charges reorganize to minimize the presence of counterions within the constraints of a given lattice. Within this framework, grafting density acts as a tunable electrostatic valence. By varying core curvature and polymer MW, we control both the magnitude and asymmetry of surface charge density while maintaining a constant pH. When the charge balance is nearly symmetric, parent 1:1 lattices such as ZnS or CsCl are stabilized. Increasing charge asymmetry frustrates strict $AB$ neutrality, promoting progressive occupation of interstitial sites or reorganization into larger unit cells, thereby stabilizing higher-stoichiometry phases such as \ch{CaF2} and \ch{Th3P4}. A detailed discussion on NP neutrality is in SI Section~9.

To rationalize this observation, we define the asymmetry parameter

\begin{equation}
\kappa=\frac{\max\!\left(N_{\rm lig}^{\rm COOH},\,N_{\rm lig}^{\rm NH_2}\right)}
{\min\!\left(N_{\rm lig}^{\rm COOH},\,N_{\rm lig}^{\rm NH_2}\right)}\,,
\end{equation}
where $N_{\rm lig}^{\rm COOH}$ and $N_{\rm lig}^{\rm NH_2}$ are the numbers of grafted ligands on the \ch{COOH}- and \ch{NH2}-terminated NPs, respectively, as determined experimentally by thermogravimetric analysis (TGA) and summarized in SI Section~9 and Table~S9. Larger $\kappa$ indicates stronger charge-asymmetry and, empirically, facilitates access to stoichiometries beyond 1:1. We denote each binary NP pair as $(n,m)-(p,q)$, where $(n,m)$ specifies a \ch{COOH} particle (\ch{COOH-PEG}(\textit{m})\text{k--Au}(\textit{n})) with Au core diameter $n$~nm and PEG MW $m$~kDa, and $(p,q)$ specifies the corresponding \ch{NH2} particle (\ch{NH2-PEG}(\textit{q})\text{k--Au}(\textit{p})). The derived $\kappa$ values are summarized in Table \ref{Table:zeta}, with a detailed discussion provided in SI Section~8.

\vspace{-0.5 cm}

\begin{center}
\renewcommand{\arraystretch}{1.2}
\setlength{\tabcolsep}{8pt}
\begin{table}[h]
\begin{tabular}{l c c l c}
\hline
\textbf{Parent} & $\boldsymbol{\kappa}$ &  & \textbf{Derived} & $\boldsymbol{\kappa}$\\
\hline
ZnS     & 3.7 & $\rightarrow$ & \ch{CaF2}   & 4.4 \\
\multicolumn{2}{l}{\footnotesize $(5,5)-(10,10)$} & & \multicolumn{2}{l}{\footnotesize $(5,2)-(10,5)$} \\
\hline
Diamond & 1.1 & $\rightarrow$ & \ch{A3}     & 1.6 \\
\multicolumn{2}{l}{\footnotesize $(10,10)-(10,5)$} & & \multicolumn{2}{l}{\footnotesize $(10,2)-(10,5)$} \\
\hline
CsCl    & 4.1$^{\ast}$ & $\rightarrow$ & \ch{Th3P4}  & 4.1 \\
\multicolumn{2}{l}{\footnotesize $(5,5)-(10,5)$} & & \multicolumn{2}{l}{\footnotesize $(10,5)-(5,5)$} \\
\hline
\end{tabular}
\caption{Examples of how stoichiometry is achieved by inducing grafting density asymmetry, as parameterized by $\kappa$. The  positive(-NH$_2$) is less effective than the negative(-COOH) so a (10,5)-(5,5) is more asymmetric than the  analogous (5,5)-(10,5), which is what $\ast$ indicates.}\label{Table:zeta}
\end{table}
\end{center}
\vspace{-1 cm}

Intermediate states observed during the ZnS–\ch{CaF2} transition further support the charge-neutrality picture: partial interstitial occupation reflects incomplete compensation of the local electrostatic imbalance. Because free-charge NPs incur a substantial entropic and electrostatic penalty, assembly into higher-stoichiometry lattices reduces the population of unpaired particles in solution, lowering the overall free energy.
An important contrast is provided by the NaCl lattice. Unlike ZnS or CsCl, the rock-salt structure is relatively dense and does not contain symmetry-related interstitial sublattices that can be progressively populated. As a result, extending NaCl-type assemblies to higher $A_mB_n$ stoichiometries through systematic site filling is inherently constrained. This highlights that access to hierarchical stoichiometries requires a parent framework with available symmetry-permitted sites or a structural motif that can reorganize into a larger basis. In our case, we observe stoichiometric expansion from ZnS and CsCl into higher-order phases, but not from NaCl. Notably, this electrostatic mechanism stabilizes intrinsically open superlattices without directional bonding or rigid templates. The resulting structures arise from the interplay between long-range Coulomb correlations and excluded-volume packing, demonstrating that polymer-mediated charge regulation provides a general route to hierarchical $A_mB_n$ NP crystals beyond unit stoichiometry.

% , suggesting that the denser NaCl framework provides fewer pathways for accommodating additional particles.

The other relevant parameter is $\gamma$. In the limit of very small NPs, the ideal $\gamma_{c}$ values accurately predict the expected transitions, see Fig.~\ref{fig:schematics}c). The CaF$_2$ superlattice has $\gamma_c=\frac{1}{2}\sqrt{6}-1\approx 0.224745$, so it is stable up to a very large asymmetry. Note that in our experiments, CaF$_2$ is stable for $\gamma=0.7-0.8$, close to the $\gamma$ value of the molecular lattice itself. Notably, the packing fraction for A$_3$ is given as $\phi_{A_3}=\frac{3 \sqrt{3}}{32}\pi\approx 0.510131$, which is slightly lower than simple cubic $\phi_{sc}=\frac{\pi}{6}\approx 0.523599$, thus demonstrating the stability of open structures through the long range electrostatic force.

The Th$_3$P$_4$ structure has $\gamma_c=\frac{\sqrt{966}}{21}-1\approx 0.480026$, with the experimental value of $\gamma=0.8-0.98$, confirming the shift of larger NPs towards stability to larger $\gamma$ values.  The highest packing fraction
\begin{equation}
    \phi(\gamma=\gamma_c)=\frac{872
    \sqrt{69}-1719\sqrt{14}}{3456} \approx 0.737644 \ ,
\end{equation}
which does not exceed the single-component close packing fraction of 
$\phi_{fcc}=\frac{\pi}{3\sqrt{2}}\approx 0.740480$. 

\section{Conclusions}

This work establishes a simple, robust electrostatic mechanism for realizing open superlattices with general stoichiometry. Using \textit{in situ} SAXS, we show that ZnS-type order emerges when charge regulation is compatible with local neutrality, whereas deliberate charge mismatch, implemented by tuning grafting density through polymer MW and NP curvature, promotes systematic occupation of tetrahedral interstitial sites and drives the formation of the fluorite superlattice. Bulk particle ratio provides an independent control parameter that modulates crystallinity and suppresses excess background, confirming that stoichiometric balance in suspension is essential for fully realizing higher-stoichiometry phases. 

Beyond interstitial filling, we identify a second hierarchical pathway to complex stoichiometry: cluster reorganization. In this case, the parent \ch{CsCl} motif is preserved locally but reorganized into the larger 28-particle basis of the \ch{Th3P4} lattice. Thus, higher stoichiometry can emerge either through progressive site occupation (ZnS to \ch{CaF2}) or through motif clustering that partially retains local coordination while expanding the unit cell (\ch{CsCl} to \ch{Th3P4}).
A companion study by Sbalbi et al. reports closely related, independently obtained results using a complementary experimental strategy. In their work, interactions among the grafted AuNPs are tuned in situ via salt-mediated control of effective valency, leading to the crystallization of CsCl-, \ch{Th3P4}-, \ch{CaF2}-, and NaCl-type lattices. The convergence of these approaches highlights the central role of tunable charge and effective valency in directing nanoparticle superlattice formation.

We further identify NP \ch{A3} and \ch{A7} superlattices with no atomic analogue, underscoring that polymer coronas function as soft frameworks that define coordination sites and enable emergent symmetries inaccessible to atomic crystals. The significance of this result is reinforced by recent theoretical predictions of a broad family of photonic band-gap superstructures, including the exotic \ch{A3} lattice reported here.\cite{cersonsky_diversity_2021} Furthermore, temperature acts as a reversible actuator: heating enhances structural order and produces pronounced negative thermal expansion through corona contraction. The improved structural integrity of ZnS and NaCl superlattices upon NP upscaling offers a practical route to defect-suppressed diamond and simple-cubic superstructures for photonic applications. Together, these findings establish a general design framework for programming stoichiometry, lattice symmetry, and responsive mechanics in charge-regulated NP crystals.

\section{Methods}

\subsection{Materials}
Citrate-stabilized AuNPs with nominal core diameters of 5, 10, and 20 nm were purchased from Ted Pella Inc. Particle sizes and dispersities were independently verified by Transmission electron microscopy (TEM; see Figs. S2 -- S4) and SAXS.\cite{nayak2023tuning} PEG ligands with number-average MWs of 2, 5, and 10 kDa, functionalized with a thiol (\ch{-SH}) group at one terminus and either a carboxyl (\ch{-COOH}) or amine (\ch{-NH2}) group at the other, were purchased from Creative PEGworks (USA) and used as received. Aqueous solutions were prepared using ultrapure deionized water (Milli-Q, resistivity 18.2 M$\Omega$·cm at 25~\textdegree C). Solution pH was adjusted using hydrochloric acid (HCl, Fisher Scientific), which was used without further purification.

\vspace{-0.4 cm}
\subsection{PEG Grafting to AuNPs Surfaces}
AuNPs were functionalized with HS–PEG\-–COOH or HS–PEG–\ch{NH2} ligands using a ligand-exchange protocol, following established procedures.\cite{nayak2025effect,kim2023two} PEG ligands were first dissolved in ultrapure water to form homogeneous solutions and then added in excess of molars to aqueous AuNP suspensions. The AuNPs-to-PEG chain ratio was set to 1:1500 for 5 nm AuNPs, 1:6000 for 10 nm AuNPs, and 1:6400 for 20 nm AuNPs. The resulting mixtures were rotated overnight at approximately 35 RPM using a Roto-Shake Genie (Scientific Industries, NY, USA) to promote efficient ligand exchange.

The excess unbound PEG was removed by three successive rounds of centrifugation. For 5 nm AuNPs, samples were centrifuged at 21,000 ×g for 90 min per cycle, while 10 nm AuNPs were centrifuged at 20,000 ×g for 75 min per cycle and 12000 ×g for 20 nm AuNPs. After each centrifugation step, the supernatant was discarded and the PEG-functionalized AuNPs were resuspended in Milli-Q water to obtain stable stock suspensions.

The concentration of PEG-grafted AuNPs was determined using ultraviolet–visible absorption spectroscopy (UV–vis) (NanoDrop One Microvolume, Thermo Fisher Scientific). Stock suspensions were adjusted to final concentrations of approximately 80 nM for 5 nm AuNPs, 20 nM for 10 nm AuNPs, and 4 nM for 20 nm AuNPs. AuNP molarities were calculated based on particle number concentrations provided by the manufacturer (Ted Pella Inc.) and are described in previous publication.\cite {nayak2026valence,nayak_assembling_2023}

\vspace{-0.5 cm}
\subsection{DLS and $\zeta-$potential Measurements}
The hydrodynamic diameters ($D_{\mathrm{H}}$) of PEG-grafted AuNPs were measured by dynamic light scattering (DLS) using a NanoZS90 instrument equipped with Zetasizer software (Malvern, UK). $\zeta$-potential measurements were performed on the same instrument to quantify the effective surface charge of the functionalized NPs. The resulting $D_{\mathrm{H}}$ and $\zeta$-potential values are shown in Fig. S1 and summarized in Table~S1.

Successful PEG grafting and the presence of distinct terminal functionalities (\ch{-COOH} or \ch{-NH2}) were confirmed by the simultaneous increase in hydrodynamic size and the emergence of opposite surface charges. In particular, \ch{-NH2}-terminated PEG-AuNPs consistently exhibit larger $D_{\mathrm{H}}$ values than their \ch{-COOH}-terminated counterparts, an effect that becomes more pronounced at higher PEG MWs (e.g., 10 kDa). This trend reflects the more hydrated and expanded polymer corona associated with amine-terminated PEG, in contrast to the relatively compact conformation of carboxyl-terminated PEG. Additional details regarding PEG grafting and NP characterization can be found in SI Section 1.

\vspace{-0.4 cm}
\subsection{Surface Charge Manipulation}

The surface charge of PEG-grafted AuNPs was tuned by controlled acidification using HCl. Stock HCl solutions were added to the NP suspensions to achieve a final concentration of 1~mM, corresponding to a target pH of $\sim$3. Because the small sample volumes required for SAXS measurements preclude direct pH measurement, the pH values reported throughout this work are calculated from the amount of HCl added and are indicated in parentheses. These calculated values were independently validated by bulk titration of representative samples, yielding agreement within $\pm$0.2 pH units.

For SAXS experiments, calculated volumes of PEG-AuNP stock suspensions were combined in glass scintillation vials at the desired mixing ratios, defined in terms of NP number ratios. The appropriate volume of HCl stock solution was then added, followed by thorough mixing. As a representative example, approximately 22~$\mu$L of 10 mM HCl was added to a 200~$\mu$L NP mixture to achieve 1 mM HCl ($\sim$pH~3). The samples were incubated for $\sim$20~minutes to allow equilibration of surface charge prior to measurement. The resulting suspensions were then loaded into borosilicate capillaries (inner diameter $\sim$1.5~mm) for SAXS data collection.

\subsection{SAXS Measurements}

\textit{In situ} SAXS was employed to determine the structures formed in suspension as a function of pH and temperature. Owing to the substantially lower ED of the polymer corona relative to gold, the scattering signal is dominated by the AuNP cores. Binary mixtures comprising AuNPs of two distinct core sizes (5 and 10~nm) were used to provide sufficient scattering contrast between the two components. In contrast, when the binary system consisted of particles with identical core sizes (either 5 or 10~nm), the resulting superlattices exhibited a single–particle–like scattering signature, consistent with uniform lattices such as diamond or simple cubic structures.  For generality, we denote the two NP species as $\mathcal{A}$ and $\mathcal{B}$, with $D_{\mathcal{A}} > D_{\mathcal{B}}$, without explicitly specifying their core sizes or PEG terminal functionalities.

Synchrotron-based \textit{in situ} SAXS experiments were performed at beamline 12-ID-B of the Advanced Photon Source (APS), Argonne National Laboratory, and at the 11-BM CMS beamline of the National Synchrotron Light Source II (NSLS-II), Brookhaven National Laboratory. The incident X-ray energies were 13.3~keV ($\lambda \simeq 0.0932$ nm) and 13.5~keV ($\lambda \simeq 0.0918$ nm), respectively, and scattering patterns were collected using a Pilatus 2M detector. Samples were loaded into borosilicate-glass capillaries ($\sim$1.5~mm inner diameter) and mounted vertically in beamline-specific sample holders, normal to the incident X-ray beam. Temperature-dependent SAXS measurements were conducted under controlled heating and cooling conditions. At APS, sample temperature was regulated using a calibrated hot-air blower, whereas at NSLS-II temperature control was achieved using a resistive heating stage integrated into the sample environment. Cooling at both beamlines was performed using a controlled coolant flow.
Temperature was monitored and controlled using a resistance temperature detector with a tolerance of 10\%.
Temperature was ramped at a rate of 5~\textdegree C~min$^{-1}$ to the target value, followed by an equilibration period of 10~minutes prior to data acquisition. SAXS patterns were then collected at each temperature point to ensure thermal stability during measurement.

Scattering data were reduced to one-dimensional intensity profiles as a function of the scattering vector magnitude $Q = 4\pi \sin\theta / \lambda$, where $2\theta$ is the scattering angle and $\lambda$ is the X-ray wavelength. Data reduction and analysis followed standard beamline-specific protocols. Further details regarding experimental setup, acquisition procedures, and data analysis can be found in previous reports.\cite{nayak2023assembling,kim2020temperaturenanorods,morozova_colloidal_2023}

\subsection{SAXS Data Analysis}

The scattering profiles, $I(Q)$, were background-corrected by subtracting the signal from corresponding water samples, yielding diffraction patterns characteristic of polycrystalline assemblies in suspension. All scattering curves shown are presented as background-subtracted raw $I(Q)$; no form-factor correction $P(Q)$ was applied, as the scattering is dominated by long-range structural correlations. Structural assignments are therefore based primarily on peak positions, with relative intensities used as supporting, but not sole, criteria.

Data analysis followed standard approaches for X-ray diffraction from polycrystalline materials. For crystalline samples, Bragg peak positions were identified based on expected reflections for candidate space groups, and corresponding structure factors were calculated and fitted to the experimental data to extract the fundamental unit-cell parameters and particle occupancies. Structural assignments were not based solely on peak positions, but on the combined agreement of Bragg peak positions, relative intensities, and line shapes. Alternative phase-coexistence models were also considered; however, they did not reproduce the full diffraction pattern, either because they introduced additional reflections, unresolved peak overlap, or intensity ratios inconsistent with the experimental data (see SI Section~4.7).

We note that \textit{ex-situ} electron microscopy (EM) of these PEG-stabilized superlattices is technically challenging because removal from aqueous suspensions alters the pH, ionic environment, and polymer conformation that stabilize the lattice. As a result, reliable real-space imaging will likely require dedicated fixation strategies analogous to those developed for similar systems.\cite{macfarlane2011nanoparticle, Liu2016a} Consequently, structural assignments in this work are based on in situ SAXS measurements performed under native assembly conditions. A detailed discussion on the current limitations of EM is provided in SI Section 2.1.

Based on the extent of structural order, assemblies were classified into three regimes. LRO is defined by resolution-limited diffraction peaks, typically corresponding to crystalline domains exceeding $\sim$10 unit cells. SRO is assigned when peak widths are at least twice the instrumental resolution, consistent with crystalline domains smaller than $\sim$5 unit cells. Intermediate cases, characterized by domain sizes of approximately 5--10 unit cells, are referred to as MRO~\cite{nayak2023ionic,nayak2026valence}. All SAXS data, relative lattice parameters, and structure assignments are
summarized in SI Section 3 and Figs. S5–-S20.
Details of the structure-factor analysis are provided in SI Section 4 and Tables S2--S8. The \ch{Th3P4} structure calculations are described in
SI Section 6 and Figs. S21--S24.

\subsection{Theoretical considerations}
Theoretical details of the nanoparticle-neutrality framework, charge regulation, and calculation of the asymmetry parameter $\kappa$
are presented in SI Section 9, Figs. S25–-S26,
and Table S9. Additional details on packing 
fraction calculations are provided in SI Section 5.

\vspace{-0.5 cm}
\section{Data Availability}
All data supporting the findings of this study are available within the Article and its Supplementary Information. Raw datasets, including SAXS measurements, electron microscopy images, and DLS data are deposited at the Harvard Dataverse under the accession DOI: doi.org/10.7910/DVN/LIQVJC

\vspace{-0.5 cm}
\section{Acknowledgements}
The authors thank Dr. Byeongdu Lee (APS), Dr. Robert J. Macfarlane, and Mr. Nicholas Sbalbi (MIT), for helpful discussions. This work was supported by the U.S. Department of Energy (DOE), Office of Science, Basic Energy Sciences, Materials Science, and Engineering Division. The research was performed at the Ames National Laboratory, which is operated for the U.S. DOE by Iowa State University under contract No. DE-AC02-07CH11358. The authors thank the 12-ID-B beamline staff team for the synchrotron beamline support at the APS, Argonne National Laboratory. Part of this research was performed on APS beam time awards (DOI: doi.org/10.46936/APS-190536/60014439; doi.org/10.46936/APS-184180/60011855) from the Advanced Photon Source, a U.S. DOE Office of Science user facility operated for the DOE Office of Science by Argonne National Laboratory under Contract No. DE-AC02-06CH11357. Part of this research used the Complex Materials Scattering (CMS) Beamline (Beamline 11-BM) of the National Synchrotron Light Source II, which is a U.S. DOE Office of Science User Facility, operated for the DOE Office of Science by Brookhaven National Laboratory under Contract No. DE-SC0012704.

% \newpage
\vspace{-0.5 cm}
\section{Author contributions}
DV, AT, WW, and SM conceived and supervised the project. BN, PK, and WK prepared the samples. BN, PK, WW, and DV conducted the experiments at synchrotron sources. BN, WW, and DV designed the experiment and analyzed the data. AT developed the theoretical and computational models. HZ assisted in conducting and processing SAXS measurements at NSLS-II. BN, AT, and DV wrote the manuscript. SM, DV, AT, and WW acquired project funding. All authors discussed the results and reviewed the manuscript.

\vspace{-0.5 cm}
\section{Competing interests}
The authors declare that they have no competing interests.
 
\vspace{-0.5 cm}
\section{Supplementary Information}\label{SI}
The supplementary information PDF file includes the following information:
\begin{itemize}[itemsep=0.1cm, parsep=0pt]
    \item Supplementary Text: Additional DLS data, TEM Data, SAXS Data, Extended SAXS analysis, and theoretical considerations
\end{itemize}

\vspace{-0.5 cm}
\normalem
\bibliography{Ref.bib,references.bib}

\clearpage
\onecolumn

\section*{Table of Contents Graphics}

\textbf{Table of contents text:} Polymer-mediated charge regulation drives two hierarchical routes to higher-stoichiometry nanoparticle superlattices. Tuning the polymer's molecular weight promotes progressive interstitial filling, transforming ZnS into the fully occupied \ch{CaF2} lattice. Exchanging polymer assignment between large and small nanoparticles instead breaks CsCl symmetry and reorganizes the parent motif into the higher-basis \ch{Th3P4} superstructure with 28 particles per unit cell.

\vspace{1em}

\begin{center}
    \includegraphics[width=0.9\linewidth]{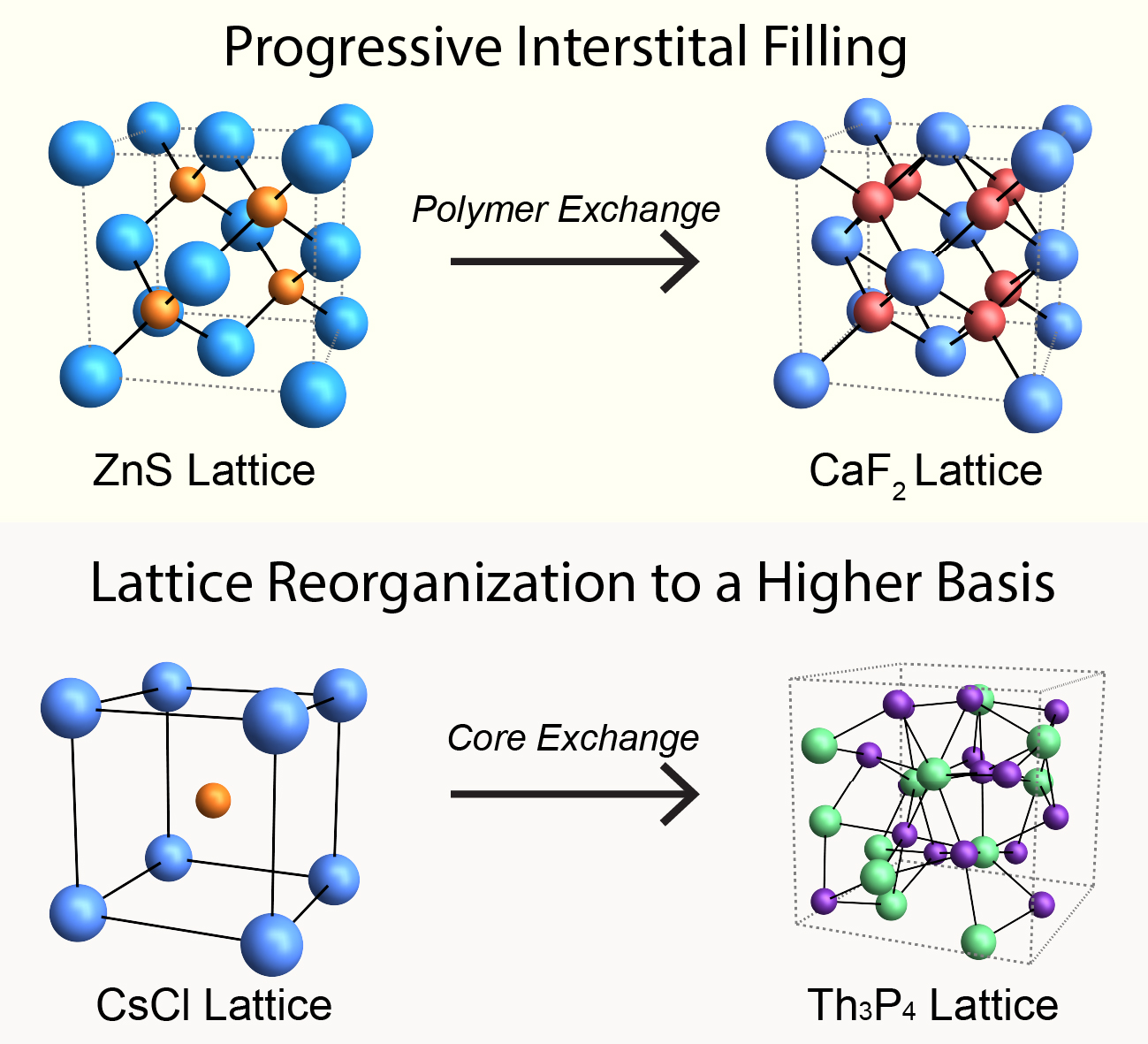}
\end{center}

\end{document}